\documentclass[11pt,epsf]{article}
\usepackage{amsmath}
\usepackage{amsfonts}
\usepackage{amssymb}
\usepackage{graphicx}
\newtheorem{theorem}{Theorem}
\newtheorem{lemma}{Lemma}
\newcommand {\dfn} {\stackrel{\Delta} {=}}

\newcommand{\eqa}{\stackrel{\mbox{(a)}}{=}}
\newcommand{\eqb}{\stackrel{\mbox{(b)}}{=}}
\newcommand{\eqc}{\stackrel{\mbox{(c)}}{=}}
\newcommand{\eqd}{\stackrel{\mbox{(d)}}{=}}
\newcommand{\eqe}{\stackrel{\mbox{(e)}}{=}}
\newcommand{\eqf}{\stackrel{\mbox{(f)}}{=}}

\newcommand{\eqh}{\stackrel{\mbox{(h)}}{=}}
\newcommand{\eqi}{\stackrel{\mbox{(i)}}{=}}

\newcommand{\lea}{\stackrel{\mbox{(a)}}{\le}}
\newcommand{\leb}{\stackrel{\mbox{(b)}}{\le}}
\newcommand{\lec}{\stackrel{\mbox{(c)}}{\le}}
\newcommand{\led}{\stackrel{\mbox{(d)}}{\le}}

\newcommand{\leg}{\stackrel{\mbox{(g)}}{\le}}
\newcommand{\gea}{\stackrel{\mbox{(a)}}{\ge}}
\newcommand{\geb}{\stackrel{\mbox{(b)}}{\ge}}

\newcommand{\gee}{\stackrel{\mbox{(e)}}{\ge}}
\newcommand{\gef}{\stackrel{\mbox{(f)}}{\ge}}
\newcommand{\geg}{\stackrel{\mbox{(g)}}{\ge}}
\newcommand {\reals} {{\rm I\!R}}

\newcommand {\bx} {\mbox{\boldmath $x$}}

\newcommand {\bE} {\mbox{\boldmath $E$}}

\newcommand {\bU} {\mbox{\boldmath $U$}}

\newcommand {\bX} {\mbox{\boldmath $X$}}

\newcommand{\calA}{{\cal A}}

\newcommand{\calC}{{\cal C}}

\newcommand{\calK}{{\cal K}}

\newcommand{\calT}{{\cal T}}
\newcommand{\calU}{{\cal U}}
\newcommand{\calV}{{\cal V}}
\newcommand{\calW}{{\cal W}}
\newcommand{\calX}{{\cal X}}
\newcommand{\calY}{{\cal Y}}
\newcommand{\calZ}{{\cal Z}}
\begin{document}
\thispagestyle{empty}
\title{Shannon's Secrecy System With
Informed Receivers and its Application to
Systematic Coding for Wiretapped Channels}
\author{Neri Merhav}
%\date{}
\maketitle

\begin{center}
Department of Electrical Engineering \\
Technion - Israel Institute of Technology \\
Technion City, Haifa 32000, ISRAEL \\
E--mail: {\tt merhav@ee.technion.ac.il}\\
\end{center}
\vspace{1.5\baselineskip}
\setlength{\baselineskip}{1.5\baselineskip}

\begin{abstract}
Shannon's secrecy system is studied in a setting, where both the legitimate decoder
and the wiretapper have access to side information sequences correlated to the source, but the
wiretapper receives both the coded information and the side information via channels
that are more noisy than the respective 
channels of the legitmate decoder, which in turn, also
shares a secret key with the encoder. A single--letter characterization is provided for the
achievable region in the space of five figures of merit: the equivocation at the wiretapper,
the key rate, the distortion of the source reconstruction at the legitimate receiver, the
bandwidth expansion factor of the coded channels, and the average transmission cost (generalized power).
Beyond the fact that this is an extension of earlier
studies, it also provides a framework for studying fundamental performance limits of systematic codes
in the presence of a wiretap channel. The best achievable performance of systematic codes is then compared
to that of a general code in several respects, and a few examples are given.\\

\noindent
{\bf Index Terms:} wiretap channel, encryption, Shannon's cipher system, separation theorem, systematic codes.
\end{abstract}

\clearpage
\section{Introduction}

Wyner, in his well--known paper on the wiretap channel 
\cite{Wyner75}, studied the problem of secure communication
across a degraded broadcast channel, without using a secret key,
where the legitimate receiver has access to the output of the good channel
and the wiretapper receives the output of the bad channel. In that paper, Wyner
characterized the optimum trade--off 
between reliable coding rates and the equivocation at the
wiretapper, which was defined in terms of 
the conditional entropy of the source given the
output of the bad channel, observed by the 
wire--tapper. Among other things, Wyner establised and characterized,
in the same paper, the notion of the 
{\it secrecy capacity}, which is the maximum coding rate that still
allows full secrecy, where the equivocation is equal to the (unconditional) entropy of the source,
thus rendering the information available to the wiretapper, virtually useless for learning anything
about the source. By applying good codes at rates close 
to the secrecy capacity, the channel is
fully exploited in the sense that the ``excess noise'', 
that is sufferred at the bad channel output (beyond
the noise at the good channel output), 
plays the role of securing the message with maximum efficiency.
The idea behind the construction of a good code for the 
wiretapped channel is essentially similar to the idea of binning. One
creates a relatively large code, which is reliably decodable at the legitimate receiver, and
which is thought of as an hierarchy of randomized sub--codes, each of which
being reliably decodable individually by the wiretapper. However, 
the bits that are decodable by the wiretapper are only those of the
randomization, and thus carry information that is irrelevant with regard to the source.

Throughout the three decades that have passed since \cite{Wyner75} was published, the results of that paper
have been extended in quite many directions, and we mention here only a few. 
Csisz\'ar and K\"orner \cite{CK78} have generalized Wyner's setting to a broadcast channel that is not
necessarily degraded (allowing also a common message to both receivers). 
Very shortly afterwards, Leung--Yan--Cheong
and Hellman \cite{LYCM78}, studied the Gaussian wiretap channel, 
and have shown, among other things that its secrecy capacity 
is simply the difference between the capacities of the main (legitimate) channel and the wiretap channel.
In \cite{OW85}, Ozarow and Wyner
studied another model, referred to as the type II wiretap channel, where the main communication channel
is noiseless, but the wiretapper has access to a subset of the coded bits, and optimal tradeoffs were
characterized. In \cite{Yamamoto89}, the wiretap channel 
model was extended to have two parallel broadcast channels,
connecting one encoder and one legitmate decoder, where
both channels are wiretapped by non--collaborating 
wiretappers, and again, optimum tradeoffs where given in terms
of single--letter expressions. In \cite{Yamamoto97}, 
the scope of \cite{Wyner75} was extended in two ways: First, by
allowing a secret key to be shared between the encoder 
and the legitimate receiver, and secondly, by allowing a certain
distortion in the reconstruction of the source at 
the legitimate receiver. The main coding theorem of \cite{Yamamoto97}
suggests a separation principle, which asserts that 
no asymptotic optimality is lost if the encoder, first, applies a
rate--distortion source code, then encrypts the compressed 
bits, and finally, applies a good code for the wiretap channel.
More recently, the Gaussian wiretap channel model of \cite{LYCM78} was further extended in two directions: one is 
the Gaussian multiple access wiretap channel of \cite{TY06}, and the other is Gaussian intereference wiretap channel
of \cite{Mitrpant03}, \cite{MHVL04}, where the encoder has access to the interference signal as side information,
similarly as in Costa's dirty paper channel \cite{Costa83}.

\begin{figure}[ht]
\hspace*{1cm}\input{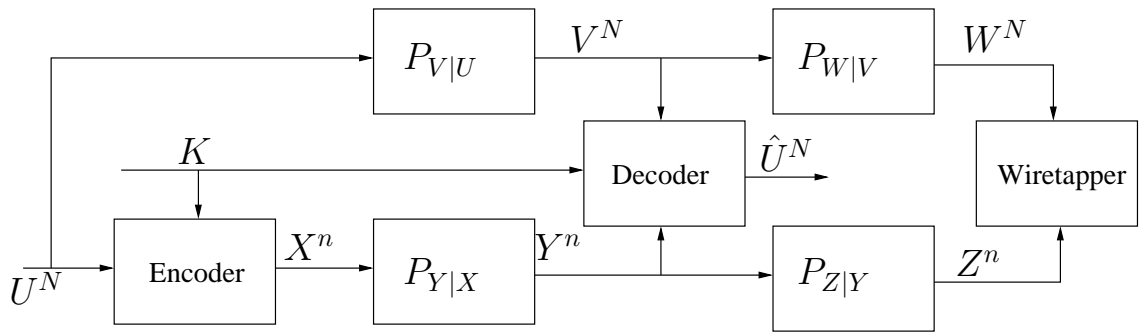}
\caption{The wiretap channel with side information at the receivers.}
\label{gen}
\end{figure}

In this paper, we extend the setting of 
the wiretap channel in a different direction.
For simplicity, we adopt the structure of a degraded 
broadcast channel, as in \cite{Wyner75} (though it is plausible
that the results are generalizable to more general broadcast channels), and
similarly as in \cite{Yamamoto97},
we allow a secret key shared between the 
encoder and the authorized decoder, as well as lossy reconstruction of the
source within a prescribed distortion level, but
we, moreover, allow also side informations, correlated to the source, to be
available both to the legitimate decoder 
and the wiretapper. We assume that 
the wiretapper receives its side information
via a channel that is degraded relative 
to the side information channel of the 
innocent decoder (see Fig.\ \ref{gen}). Our main result is a
single--letter characterization of the optimum tradeoff among
five figures of merit: the equivocation at the wiretapper, the 
distortion level in reconstructing the source at the authorized decoder,
the bandwidth expansion factor of the coded channels, the rate
of the secret key relative to the source, and the average tranmission cost.

One of the motivations for this study 
is that it establishes a framework for deriving performance 
limits of systematic codes for wiretapped channels 
and assessing their loss in performance compared to general codes
(as was done in \cite{SVZ98} in a different context): 
The side information channels 
($P_{V|U}$ and $P_{W|V}$ in Fig.\ \ref{gen}) 
can be thought of as conveying the
systematic (uncoded) part of the codeword. 
We compare 
the best achievable performance of systematic codes
to that of general codes at the same coding rates, in several aspects, 
like the maximum achievable equivocation in the absence of a secret key,
the maximum achievable equivocation in the presence of a full--rate key,
the key rate needed to achieve the maximum achievable equivocation,
and the distortion achieved when the channel is utilized at rate close to
the secrecy capacity. A few examples are given for situations
where systematic codes are as good as (and sometimes even better than)
general codes.

The outline of the remaining parts of this paper is as follows:
In Section 2, we set up the notation, formulate the problem, present the main result,
and make a few comments. In Section 3, we discuss the implications
on systematic coding, and we make comparisons with general codes,
as described in the previous paragraph. In Section 4, we prove the
converse part of the main result, and finally, in Section 5, we
prove the direct part.

\section{Problem Formulation and Main Result}

We begin by establishing some notation conventions.
Throughout this paper, scalar random
variables (RV's) will be denoted by capital
letters, their sample values will be denoted by
the respective lower case letters, and their alphabets will be denoted
by the respective calligraphic letters.
A similar convention will apply to
random vectors and their sample values,
which will be denoted with same symbols superscripted by the dimension,
or by the bold face font, if there is no room for confusion regarding the dimension.
Thus, for example, $U^N$ ($N$ -- positive integer) or $\bU$
will denote a random $N$--vector $(U_1,...,U_N)$,
and $u^N=(u_1,...,u_N)$ is a specific vector value in $\calU^N$,
the $N$--th Cartesian power of $\calU$.

Sources and channels will be denoted generically by the letter $P$,
subscripted by the name of the RV and its conditioning,
if applicable, e.g., $P_U(u)$ is the probability function of
$U$ at the point $U=u$, $P_{Y|X}(y|x)$
is the conditional probability of $Y=y$ given $X=x$, and so on.
Whenever clear from the context, these subscripts will be omitted.
Information theoretic quantities like entropies and mutual
informations will be denoted following the usual conventions
of the Information Theory literature, e.g., $H(U^N)$, $I(X^n;Y^n)$,
and so on. For single--letter
information quantities (i.e., when $n=1$ or $N=1$),
subscripts will be omitted, e.g., $H(U^1)=H(U_1)$ will
be denoted by $H(U)$,
similarly, $I(X^1;Y^1)=I(X_1;Y_1)$ will be denoted by $I(X;Y)$, and so on.
For three random variables, generically denoted 
$A$, $B$, and $C$, the notation 
$A\ominus B\ominus C$ will designate the fact that
they form, in this order, a Markov chain. The extension
of this notation to longer Markov chains will be straightforward.
The cardinality of a finite set $\calA$ will be denoted by $|\calA|$.
The notation $[a]_+$ will stand for $\max\{0,a\}$. Finally,
for $a,b\in\{0,1\}$, $a\oplus b$ will denote the modulo 2 sum (XOR) of $a$
and $b$, and for two general positive integers, $a$ and $b$, the notation
$a\oplus b$ will designate the positive integer whose binary representation
is given by the bit--wise modulo 2 sum of the corresponding bits of the
binary representations of $a$ and $b$.

We now turn to the formal description 
of the model and the problem setting.
A source $P_U$ 
generates a sequence of $N$ ($N$ -- positive integer)
independent copies, $U^N=(U_1,\ldots,U_N)$, of a finite--alphabet RV,
$U\in\calU$. At the same time, a discrete memoryless channel (DMC),
symbolized by $P_{V|U}$ generates from $U^N$,
another $N$-vector $V^N=(V_1,\ldots,V_N)$, 
with components in a finite--alphabet
$\calV$, and another DMC, denoted 
$P_{W|V}$, produces from $V^N$, yet
another $N$-vector $W^N=(W_1,\ldots,W_N)$,
with components in a finite--alphabet
$\calW$. Thus, the joint probability distribution of $(u^N,v^N,w^N)$
is given by 
$$P_{U^N}(u^N)P_{V^N|U^N}(v^N|u^N)P_{W^N|V^N}(w^N|v^N)=
\prod_{i=1}^N[P_U(u_i)P_{V|U}(v_i|u_i)P_{W|V}(w_i|v_i)].$$
At the same time and independently,
another source $P_K$, henceforth referred to as the {\it key source}, 
generates a random variable (or vector) $K$ 
taking values in a finite alphabet $\calK$.

Two additional cascaded DMC's operate at a bandwidth expansion factor
of $n/N$ channel uses per source symbol.
This means that during the time that the
source generates a block $U^N$ of $N$ symbols,
the first channel receives a block $X^n$ of $n$ channel input
symbols taking on values in a finite alphabet $\calX$, and outputs
a block $Y^n$ of $n$ channel output
symbols in a finite alphabet $\calY$, according to 
$$P_{Y^n|X^n}(y^n|x^n)=\prod_{j=1}^nP_{Y|X}(y_j|x_j),$$
whereas the second DMC
receives $Y^n$ as an input vector
and outputs
a block $Z^n$ of $n$ channel output
symbols in a finite alphabet $\calZ$, according to 
$$P_{Z^n|Y^n}(z^n|y^n)=\prod_{j=1}^nP_{Z|Y}(z_j|y_j).$$

Given $N$ and $n$, a block encoder 
is a mapping $f_{n,N}:\calU^N\times\calK\to\calX^n$, 
whose output is
$X^n=(X_1,\ldots,X_n)=f_{n,N}(U^N,K)\in\calX^n$. The channel input vector should
satisfy an average transmission cost (generalized power) constraint:
\begin{equation}
\frac{1}{n}\sum_{j=1}^n\bE\{\phi(X_j)\}\le Q,
\end{equation}
where $\phi:\calX\to\reals^+$ is the generalized power function and $Q$ is a given positive real.
The corresponding block decoder (of the authorized party)
is a mapping $g_{n,N}:\calY^n\times\calV^N\times\calK\to\hat{\calU}^N$,
whose output is
$\hat{U}^N=(\hat{U}_1,\ldots,\hat{U}_N)=g_{n,N}(Y^n,V^N,K)\in\hat{\calU}^N$, 
where $\hat{\calU}$ is the reproduction alphabet of the decoder output symbols.

Let $d:\calU\times\hat{\calU}\to \reals^+$ denote
a single--letter distortion measure between 
source symbols and reproduction symbols,
and let the distortion between the vectors, 
$u^N\in\calU^N$ and $\hat{u}^N\in\hat{\calU}^N$,
be defined additively across the corresponding components, as usual. 
Let $R_{U|V}(D)$ denote the Wyner--Ziv rate--distortion function \cite{WZ76}
of the soure $U$ with
respect to the distortion measure $d$, and a decoder side information $V$, i.e.,
$$R_{U|V}(D)=\inf [I(U;A)-I(V;A)],$$
where the infimum is over all RV's $A$ with alphabet size $|\calU|+1$, that
form a Markov chain $A\ominus U\ominus V$ 
and that satisfy
$\min_{\{\psi:\calA\times\calV\to\hat{\calU}\}} \bE\{d(U,\psi(A,V))\}\le D$.
Given the degraded broadcast channel $P_{YZ|X}(y,z|x)=P_{Y|X}(y|x)P_{Z|Y}(z|y)$, 
we will also define the function 
\begin{equation}
\Gamma(r,q)=\sup_{\{P_X:~I(X;Y)\ge r,~\bE\phi(X)\le q\}}I(X;Y|Z)=
\sup_{\{P_X:~I(X;Y)\ge r,~\bE\phi(X)\le q\}}[I(X;Y)-I(X;Z)]
\end{equation}
which is similar to Wyner's $\Gamma$ function \cite{Wyner75}, but with the additional
generalized power constraint.

An $(N,n,\lambda,D,\Delta,R,Q)$ codec is an encoder--decoder pair
with parameters $N$ and $n$, that satisfies the following requirements:
\begin{itemize}
\item [1.] The bandwidth expansion factor is $n/N\le \lambda$.
\item [2.] The expected distortion 
between the source and the reproduction satisfies
\begin{equation}
\sum_{i=1}^N \bE\{d(U_i,\hat{U}_i)\}\le ND.
\end{equation}
\item [3.] The equivocation of the message source satisfies
\begin{equation}
H(U^N|W^N,Z^n)\ge N\Delta.
\end{equation}
\item [4.] The rate of the secret key is $H(K)/N\le R$.
\item [5.] The generalized transmission power satisfies $\sum_{j=1}^n \bE\{\phi(X_j)\}\le nQ$.
\end{itemize}
A quintuple $(\lambda,D,\Delta,R,Q)$ is said to be {\it achievable}
if for every $\epsilon > 0$, there is a sufficiently large $N$ and $n$
for which 
$(N,n,\lambda+\epsilon,D+\epsilon,\Delta-\epsilon,R+\epsilon,Q+\epsilon)$ codecs exist.
The {\it achievable region} of quintuples
$\{(\lambda,D,\Delta,R,Q)\}$ is the set of all achievable
quintuples $(\lambda,D,\Delta,R,Q)$. 

The following theorem characterizes 
the region of achievable quintuples $(\lambda,D,\Delta,R,Q)$.
\begin{theorem}
A quintuple $(\lambda,D,\Delta,R,Q)$ is achievable iff 
$$\Delta \le \Delta^*(\lambda,R,D,Q)\dfn H(U|W)-\left[R_{U|V}(D)-\lambda
\Gamma\left(\frac{R_{U|V}(D)}{\lambda},Q\right)-R\right]_+.$$
\end{theorem}

\noindent
{\bf Discussion:}
A few comments are in order at this point.

As mentioned in the Introduction, Theorem 1 generalizes earlier results
reported in \cite{Wyner75}, \cite{SVZ98}, and \cite{Yamamoto97}.
The generalization relative to \cite[Theorem 1, ``Case of LDBC'']{Yamamoto97} 
is primarily in the presence of side informations at the authorized decoder as
well as the wiretapper. It should be also noted that in \cite{Yamamoto97}, there is
no full proof of the direct part, but only an intuitive argument. Here, we provide
complete proofs for both the converse part and the direct part, which are both based on
the corresponding proofs in \cite{Wyner75}, but there are a few twists that
are necessary in order to incorporate the secret key, $K$, the side informations, $V^N$ and $W^N$,
and the generalized power constraint.
For example, one of the additional ingredients in the proof of the direct part, that is not
present in the direct part of \cite{Wyner75}, is that we need to show that the key $K$ can be
estimated reliably from $U^N$, $W^N$, and $Z^n$, so that $H(K|U^N,W^N,Z^n)$ is small.

As in \cite{Yamamoto97}, Theorem 1 here 
suggests a {\it separation principle}, that guarantees no loss in
asymptotically optimum performance, if one separates source coding, encryption, and channel coding.
As will be seen in the proof of the direct part, the proposed achievability scheme consists of Wyner--Ziv
rate-distortion source coding, followed by encryption of the compressed bits, followed in turn by
good channel coding for the wiretapped channel, as in \cite{Wyner75}. As is demonstrated in \cite{Merhav06},
the separation principle does not always hold in situations that involve 
source coding, encryption, and channel coding.

A few words about the intuition behind the achievable upper bound on the equivocation, 
$\Delta^*(\lambda,R,D,Q)$: For
$R \ge R_{U|V}(D)-\lambda\Gamma(R_{U|V}(D)/\lambda,Q)$, there is enough randomness to achieve
the maximum possible secrecy of $H(U^N|W^N)=NH(U|W)$, which cannot be exceeded even if the
wiretapper did not have access to $Z^n$.
For the more interesting case where $R_{U|V}(D)> \lambda\Gamma(R_{U|V}(D)/\lambda,Q)$ (which in turn means
that $R_{U|V}(D)/\lambda$ is above the secrecy capacity), and
$R < R_{U|V}(D)-\lambda\Gamma(R_{U|V}(D)/\lambda,Q)$, we can express $\Delta^*(\lambda,R,D,Q)$ as the sum of
four terms:
$$\Delta^*(\lambda,R,D,Q)=[H(U|W)-H(U|V)]+[H(U|V)-R_{U|V}(D)]+
\lambda\Gamma\left(\frac{R_{U|V}(D)}{\lambda},Q\right)+R,$$
where we have added and subtracted $H(U|V)$. Now, the first bracketed term designates the fact that the wiretapper
has side information whose quality is lower than that of the authorized user, a fact which contributes to
the equivocation. The second bracketed term designates uncertainty due to the information loss at the source
encoder (although a general coding scheme may not necessarily use a source encoder explicitly). Out of the
$NR_{U|V}(D)$ bits of the description of the source, $NR$ bits are covered by the key and another
$n\Gamma(R_{U|V}(D)/\lambda,Q)=N\lambda\Gamma(R_{U|V}(D)/\lambda,Q)$ bits are covered by good channel coding for
the wiretapped channel, as in \cite{Wyner75}, \cite{Yamamoto97}. 
In designing a good coding scheme, it should be kept in mind then,
that there should be no overlap between the set of bits encrypted by the key and those that are ``hidden''
by coding. It is interesting to note that in the above decomposition of $\Delta^*(\lambda,R,D,Q)$, the
first term depends solely on the joint distribution of $(U,V,W)$, and not on any other factor of the problem,
the second term depends only on the joint distribution
of $(U,V)$ and the allowed distortion (but no longer on 
the joint distribution with $W$), and the third term depends also the coded channels.
Referring to the previous comment, it is interesting to note that even 
in the lossless case ($D=0$) and even 
if the coded channels are clean (i.e., $X^n=Y^n=Z^n$ with probability one),
the presence of side information at the
legitimate decoder, which is of better quality than the one at the wiretapper,
gives rise to ``inherent secrecy,'' that is present
even without a secret key. In such a case, the last three terms
in the above representation of $\Delta^*(\lambda,R,D,Q)$ all vanish, but the first term is still positive.
For example, a Slepian--Wolf encoder for a source $U$ and side information $V$, which is  
based on random binning, has the maximum achievable inherent secrecy of $I(U;V)$
bits/symbol if a wiretapper that observes the compressed bits has no side information.
This is in contrast to the case without side information, where there is no inherent secrecy at all.

An interesting question that arises 
is about optimum strategies and performance limits 
if one is interested to maximize the equivocation of $\hat{U}^N$ 
instead of, or in addition to that of $U^N$ (see also
\cite{Merhav06}), which is reasonable 
because it is $\hat{U}^N$ that is the information conveyed from the source.
In contrast to \cite{Merhav06}, where the problem was fully solved using ordinary rate--distortion coding
considerations, here, because of the presence of side information, the problem remains open.

Finally, as mentioned already in the Abstract and the Introduction, Theorem 1 provides a framework 
for studying the fundamental performance limits of
{\it systematic} (not necessarily linear)
codes, in the same manner as in \cite{SVZ98},
for the wiretap channel. The next section is devoted to 
such a study.

\section{Systematic Vs.\ Non--Systematic Codes}

If $\calU=\calX$, $\calV=\calY$, and the uncoded channel, $P_{V|U}$, is understood as an additional use
of the same physical channel as the coded 
channel, $P_{Y|X}$, and if $\bE\{\phi(U)\}\le Q$,
then the uncoded path $U^N\to V^N$ may
be thought of as corresponding to the transmission and reception of the
systematic (uncoded) part of a systematic code, where the information
symbols are sent directly to the channel. The total bandwidth expansion factor
of this systematic code, when the uncoded part is viewed as part of the code,
is then $(N+n)/N=1+\lambda$, assuming that $n/N=\lambda$. For
a fully coded (general, non--systematic) system with the same bandwidth expansion factor, we can
use the formula of $\Delta^*(\lambda,R,D,Q)$, but
replace $\lambda$ by $1+\lambda$ and eliminate the side informations, $V^N$ and $W^N$.
The resulting maximum achievable equivocation of a general code, is therefore:
\begin{equation}
\Delta_{\mbox{gen}}^*(\lambda,R,D,Q)
=H(U)-\left[R_U(D)-(1+\lambda)\Gamma\left(\frac{R_U(D)}{1+\lambda},Q\right)-R\right]_+,
\end{equation}
where $R_U(D)$ is the ordinary rate--distortion function of $U$ (without side information), and we are
interested to compare this to the 
original expression of $\Delta^*(\lambda,R,D,Q)$, given in Theorem 1,
which will be denoted by $\Delta_{\mbox{sys}}^*(\lambda,R,D,Q)$ throughout this section. Quite obviously,
$\Delta_{\mbox{sys}}^*(\lambda,R,D,Q)$ cannot exceed $\Delta_{\mbox{gen}}^*(\lambda,R,D,Q)$, but
it is interesting to identify cases of equality, simply by
comparing the two expressions.
We will, however, focus here on a few specfic aspects of comparison
between optimum systematic codes and optimum general codes:
\begin{enumerate}
\item The {\it full equivocation}, that is, 
the maximum equivocation that can be achieved in the
absence of limitations on the key rate (in which case, the bracketed term of $\Delta^*$ vanishes).
\item The {\it zero key--rate
equivocation}, which is defined as $\Delta^*$ for $R=0$. This quantity manifests the ``inherent'' security
that is already present in the system even without a key.
It should be noted that whenever 
$\Delta_{\mbox{sys}}^*(\lambda,0,D,Q)=\Delta_{\mbox{gen}}^*(\lambda,0,D,Q)$,
then, in general (as can be seen from the expressions of 
$\Delta_{\mbox{sys}}^*(\lambda,R,D,Q)$ and $\Delta_{\mbox{gen}}^*(\lambda,R,D,Q)$),
there is a range of $R$, 
where $\Delta_{\mbox{sys}}^*(\lambda,R,D,Q)=\Delta_{\mbox{gen}}^*(\lambda,R,D,Q)$
since, in that range, both $\Delta_{\mbox{sys}}^*(\lambda,R,D,Q)$ and 
$\Delta_{\mbox{gen}}^*(\lambda,R,D,Q)$ grow linearly with a slope of 45 degrees, starting from their 
respective values at $R=0$.
\item The {\it saturation key rate}, which is the smallest value of $R$, for which $\Delta^*$
achieves the full equivocation. When the saturation key rate is small, then so are the randomization resources
required.
\item The {\it secrecy distortion}, 
which is the value of $D$ for which the channel coding rate
equals the secrecy capacity, in other words, 
the first argument of the function $\Gamma$ agrees with the
secrecy capacity. 
This is an interesting working point, 
because it is the point where the full
equivocation is achieved without using a key at all. 
In other words, using the terminology 
that we have already defined, the zero key--rate equivocation is
equal to the full equivocation, and the saturation key rate vanishes.
\end{enumerate}

While under the first two criteria, 
systematic codes can never be strictly better than general
codes, this is not necessarily the case with the last two criteria,
because codes that are optimum in the maximum equivocation sense may be
suboptimal under other criteria.
We next compare optimum systematic codes to 
optimum codes from the above four aspects.

\vspace{0.5cm}

\noindent
{\bf 1. The full equivocation:}
Obviously, this quantity is $H(U|W)$ for 
systematic codes and $H(U)$ for general codes, thus the difference,
$I(U;W)$, depends only on the joint distribution of $U$ and $W$. In this 
respect, optimum systematic codes are
as good as optimum general codes only 
if the side information $W$ is independent of $U$ and hence useless.

\vspace{0.5cm}

\noindent
{\bf 2. The zero key--rate
equivocation:}
For $R=0$, we have
\begin{equation}
\Delta_{\mbox{gen}}^*(\lambda,0,D,Q)
=H(U)-\left[R_U(D)-
(1+\lambda)\Gamma\left(\frac{R_U(D)}{1+\lambda},Q\right)\right]_+
\end{equation}
for general codes, and
\begin{equation}
\Delta_{\mbox{sys}}^*(\lambda,0,D,Q)
=H(U|W)-\left[R_{U|V}(D)-
\lambda\Gamma\left(\frac{R_{U|V}(D)}{\lambda},Q\right)\right]_+
\end{equation}
for systematic codes. Let us assume that 
the bracketed terms in both expressions are positive
(otherwise, we are back to the comparison 
of the previous paragraph). 
Comparing the two expressions, we see that
equality is achieved if
\begin{equation}
\label{equality}
(1+\lambda)\Gamma\left(\frac{R_U(D)}{1+\lambda},Q\right)-
\lambda\Gamma\left(\frac{R_{U|V}(D)}{\lambda},Q\right)=
R_U(D)-R_{U|V}(D)-I(U;W).
\end{equation}
As is shown in \cite[eqs.\ (2.12), (2.13)]{SVZ98}, 
the difference $R_U(D)-R_{U|V}(D)$
is never larger than $I(U;V)$, but 
there are cases of equality, most notably, the lossless case $D=0$, as
$R_U(0)=H(U)$ and $R_{U|V}(0)=H(U|V)$.\footnote{Another example is the Gaussian source $U$,
the Gaussian channel $P_{V|U}$, and the squared error distortion measure, 
where $R_U(D)-R_{U|V}(D)=[h(U)-\frac{1}{2}\log(2\pi e D)]-[h(U|V)-\frac{1}{2}\log(2\pi e D)]=I(U;V)$ 
throughout the entire
interesting range of distortion levels.}
Thus, at least in the lossless case, eq.\ (\ref{equality}) boils down to
\begin{equation}
\label{equality0}
(1+\lambda)\Gamma\left(\frac{H(U)}{1+\lambda},Q\right)-
\lambda\Gamma\left(\frac{H(U|V)}{\lambda},Q\right)=
I(U;V)-I(U;W).
\end{equation}
Now, in quite a few examples of interest,
$\Gamma(r,q)$ is 
equal to a constant, $\Gamma_0$, 
throughout the entire interesting range of $r$. 
One such example occurs when $q=\infty$ (i.e., no generalized power constraint), $P_{Y|X}$ 
is the noiseless binary channel and $P_{Z|Y}$ is a binary
symmetric channel (BSC) with 
crossover probability $p_0$ (cf.\ \cite[p.\ 1362]{Wyner75}), 
in which case, $\Gamma_0=h(p_0)$, where 
$h(\cdot)$ is the binary entropy function. In this case,
the left--hand side of eq.\ (\ref{equality0}) 
becomes $h(p_0)$ independently of $\lambda$.
Now, if  $P_{V|U}$ has the same
characteristics as $P_{Y|X}$, and similatry $P_{W|V}$ has the same
characteristics as $P_{Z|Y}$ (which 
is indeed the case in systematic coding applications), and if $P_U$
is the binary symmetric source (BSS), then it
achieves the maximum of $I(U;V)-I(U;W)$, which is, again, $\Gamma_0=h(p_0)$. 
In this case, therefore, the equality 
(\ref{equality0}) is achieved. 
Similarly, if $P_{Y|X}$ is noiseless as before, 
but $P_{Z|Y}$ is an erasure channel with
erasure probability $p_0$, then 
$\Gamma_0=p_0$, and once again, equality is achieved if
$P_U$ is the BSS. Yet another example 
of this type occurs when both $P_{Y|X}$ and $P_{Z|Y}$ are
independent Gaussian channels with an input power constraint defined in terms of $\phi(x)=x^2$
(and hence, so are $P_{V|U}$ and $P_{W|V}$). 
In this case, as was shown in \cite{LYCM78},
$\Gamma_0=C_{X\to Y}-C_{X\to Z}$, 
the difference between the capacities of the channels $P_{Y|X}$ and
$P_{Z|X}$. Here, equality in (\ref{equality0}) 
is achieved if $U$ is a zero--mean Gaussian random variable
whose variance coincides with the maximum allowable input power, $Q$.
Thus, we have demonstrated a few
non--trivial examples where optimum systematic codes are 
as good as optimum codes in the absence of a secret key.

\vspace{0.5cm}

\noindent
{\bf 3. The saturation key rate:}
Here, we obtain
\begin{equation}
R_{\mbox{gen}}^*=R_U(D)-(1+\lambda)\Gamma\left(\frac{R_U(D)}{1+\lambda},Q\right)
\end{equation}
for general codes, and
\begin{equation}
R_{\mbox{sys}}^*=R_{U|V}(D)-\lambda\Gamma\left(\frac{R_{U|V}(D)}{\lambda},Q\right)
\end{equation}
for systematic codes. 
The condition for having a smaller saturation key rate for systematic codes is
\begin{equation}
\label{equality1}
(1+\lambda)\Gamma\left(\frac{R_U(D)}{1+\lambda},Q\right)-
\lambda\Gamma\left(\frac{R_{U|V}(D)}{\lambda},Q\right)\le
R_U(D)-R_{U|V}(D),
\end{equation}
namely, the comparison is similar to the 
one made with regard to the zero key--rate equivocation criterion, 
but without the term $I(U;W)$. As we have previously shown examples of
equality, even in the presence of the term $I(U;W)$, 
then the same examples can serve now for the desired inequality
in the absence of this term. 
In these examples, as well as in many others, optimum systematic codes
are advantageous over optimum codes in general.

\vspace{0.5cm}

\noindent
{\bf 4. The secrecy distortion:} As mentioned earlier, Wyner \cite{Wyner75} 
has established the notion of the secrecy capacity, $C_s$,
which is the maximum coding rate for which full secrecy 
is still achieved even without a key. 
Here we ask
how do systematic-- and non--systematic codes compare in terms of
the distortion, $D$, for which the rate of 
the channel code meets the secrecy capacity.
For non--systematic codes, this distortion level is given by the solution
to the equation 
\begin{equation}
\frac{R_U(D)}{1+\lambda}=C_s,
\end{equation}
which is 
\begin{equation}
D_{\mbox{gen}}^*=D_U((1+\lambda)C_s),
\end{equation}
where $D_U(\cdot)$
is the ordinary distortion--rate function of $U$ (without side information).
For systematic coding, on the other hand, it is the solution to the equation
\begin{equation}
\frac{R_{U|V}(D)}{\lambda}=C_s,
\end{equation}
which is 
\begin{equation}
D_{\mbox{sys}}^*=D_{U|V}(\lambda C_s),
\end{equation}
where $D_{U|V}(\cdot)$
is the Wyner--Ziv distortion--rate function of $U$ with side information $V$.
The answer to the question: which class of 
codes is better in terms of the secrecy distortion,
depends on the parameters of the problem. 
One simple extreme example pertains to the case $C_s=0$ 
(which happens, e.g., when the channel $P_{Z|Y}$ is clean and hence $Z^n=Y^n$ with probability one).
In this case,
\begin{equation}
D_{\mbox{sys}}^*=D_{U|V}(0)=\min_{\{\psi:\calV\to\hat{\calU}\}}\bE\{d(U,\psi(V))\}
\end{equation}
is clearly smaller than
\begin{equation}
D_{\mbox{gen}}^*=D_{U}(0)=\min_{\hat{u}\in\hat{\calU}}\bE\{d(U,\hat{u})\}.
\end{equation}
While the case 
where $C_s$ is strictly zero, clearly trivializes the whole problem altogether,
it is, of course, conceivable that for small enough positive values of $C_s$, 
continuity arguments imply that systematic codes
still outperform non--systematic codes in the secrecy distortion sense.

As a somewhat less trivial 
example, consider the case where $U$ is zero--mean, Gaussian,
with variance $\sigma_U^2$, the channels are Gaussian and independent, and 
$d$ is the squared error criterion. Then,
\begin{equation}
D_{\mbox{gen}}^*=\sigma_U^2\cdot 2^{-2(1+\lambda)(C_{X\to Y}-C_{X\to Z})},
\end{equation}
whereas 
\begin{equation}
D_{\mbox{sys}}^*=\sigma_{U|V}^2\cdot 2^{-2\lambda(C_{X\to Y}-C_{X\to Z})},
\end{equation}
where $\sigma_{U|V}^2$ is the minimum mean squared error associated
with optimum (linear) estimation of $U$ based on $V$.
Thus, $D_{\mbox{sys}}^* \le D_{\mbox{gen}}^*$ whenever 
\begin{equation}
C_{X\to Y}-C_{X\to Z} \le \frac{1}{2}\log\frac{\sigma_U^2}{\sigma_{U|V}^2}
=\frac{1}{2}\log\left(1+\frac{\sigma_U^2}{\sigma^2}\right),
\end{equation}
where $\sigma^2$ is the variance of the 
noise of the (Gaussian) channel from $U$ to $V$.
Note that the dependence upon $\lambda$ disappeared.
The last inequality is clearly met if, 
for example, the channel $P_{Y|X}$ is the same as the
channel $P_{V|U}$ and $\sigma_U^2=Q$ 
(in which case, the right--hand side becomes $C_{X\to Y}$). 

Note that in this aspect of the secrecy distortion, our comparison
between systematic codes and non--systematic 
codes is of the same spirit as in \cite{SVZ98},
in the sense that both are about equating rate--distortion
functions to capacities.
The only difference is that here, as opposed to \cite{SVZ98}, 
$C_s$ replaces $C_{X\to Y}$ 
in the these equations (as there is only
one coded channel and one uncoded channel in \cite{SVZ98}). 
Obviously, in the comparisons carried out in \cite{SVZ98}, systematic codes
can never outperform non--systematic codes. By contrast, as we have seen here, 
when the secrecy capacity is
the working point, this becomes possible.

\vspace{0.5cm}

\noindent
Finally, one more comment is in order regarding systematic codes:
In a real systematic code for the wiretap channel,
there is, in principle, the freedom to use part of the secret key in order to
encrypt the systematic symbols as well. 
This freedom has not been exploited thus far, and
the question is whether there is any advantage in doing so.
Suppose that the source $U$ is binary 
and the key rate is $R$ bits per source symbol ($R\le 1$).
Consider the following coding scheme. 
We select $0\le R' \le R$, and for each block $U^N$, we
use $NR'$ key bits to encrypt the 
systematic part and $N(R-R')$ key bits to encrypt the
Wyner--Ziv rate--distortion codeword 
before it is fed into the channel encoder of \cite{Wyner75}
(see also the proof of the direct part in Section 5). 
Then, by a slight extension of
the analysis in Section 5 to follow, the resulting equivocation is essentially
\begin{equation}
\Delta\approx R'H(U)+(1-R')H(U|W)+
(R-R')+\lambda\Gamma(R_{U|V}(D)/\lambda,Q)-R_{U|V}(D).
\end{equation}
Since the coefficient of $R'$, in this expression, 
is $I(U;W)-1 < 0$, the best choice of $R'$, in this example, is $R'=0$,
namely, secret key bits should better
not be used for encrypting the systematic bits, but only the coded bits,
as we assumed thus far.

\section{Proof of the Converse Part of Theorem 1}

Let an $(N,n,\lambda+\epsilon,D+\epsilon,\Delta-\epsilon,R+\epsilon,Q+\epsilon)$ codec be given.
Consider first the following chain of inequalities, which will be used
later on.
\begin{eqnarray}
\label{2nd}
I(X^n;Y^n|K,V^N,W^N)&\gea&I(U^N;Y^n|K,V^N,W^N)\nonumber\\
&=&\sum_{i=1}^NI(U_i;Y^n|K,V^N,W^N,U^{i-1})\nonumber\\
&=&\sum_{i=1}^N[H(U_i|K,V^N,W^N,U^{i-1})-H(U_i|Y^n,K,V^N,W^N,U^{i-1})]\nonumber\\
&\geb&\sum_{i=1}^N[H(U_i|V_i)-H(U_i|Y^n,K,V^N)]\nonumber\\
&\eqc&\sum_{i=1}^N[H(U_i|V_i)-H(U_i|V_i,A_i)]\nonumber\\
&=&\sum_{i=1}^NI(U_i;A_i|V_i)\nonumber\\
&=&\sum_{i=1}^N[H(A_i|V_i)-H(A_i|U_i,V_i)]\nonumber\\
&\eqd&\sum_{i=1}^N[H(A_i|V_i)-H(A_i|U_i)]\nonumber\\
&=&\sum_{i=1}^N[I(U_i;A_i)-I(V_i;A_i)]\nonumber\\
&\gee&\sum_{i=1}^NR_{U|V}(\bE d(U_i,[g_{n,N}(A_i,V_i)]_i))\nonumber\\
&\gef&NR_{U|V}\left(\frac{1}{N}\sum_{i=1}^N\bE d(U_i,[g_{n,N}(A_i,V_i)]_i)\right)\nonumber\\
&\geg&NR_{U|V}(D+\epsilon),
\end{eqnarray}
where (a) follows from the fact that $U^N\ominus (K,V^N,W^N,X^n)\ominus Y^n$ is a Markov chain,
(b) is because conditioning reduces entropy, in (c) -- $A_i$ is defined
as $(Y^n,K,V^{i-1},V_{i+1}^N)$, (d) is because $S_i\ominus U_i\ominus V_i$ is a Markov chain,
(e) is by definition of the Wyner--Ziv rate--distortion function, where $[g_{n,N}(A_i,V_i)]_i$ is the
projection of $g_{n,N}(A_i,V_i)=g_{n,N}(Y^n,V^N,K)$ to the $i$--th component,
(f) is due to the convexity of the Wyner--Ziv 
rate--distortion function \cite{WZ76},\cite[Lemma 14.9.1, p.\ 439]{CT91},
and (g) is due to its monotonicity, and the hypothesis that the codec achieves distortion $D+\epsilon$.

We next derive two upper bounds on $\Delta$.
The first one is trivial:
\begin{equation}
\label{12}
\Delta-\epsilon\le \frac{H(U^N|W^N,Z^n)}{N}\le \frac{H(U^N|W^N)}{N}=H(U|W),
\end{equation}
and so, 
\begin{equation}
\label{1stbound}
\Delta\le H(U|W) 
\end{equation}
due to the arbitrariness of $\epsilon$.
The other, more interesting, upper bound on $\Delta$ is obtained as follows:
First, we observe that
\begin{equation}
\label{2terms}
N(\Delta-\epsilon)\le H(U^N|W^N,Z^n)
=I(U^N;V^N,K|W^N,Z^n)+H(U^N|V^N,K,W^N,Z^n).
\end{equation}
Next, we bound from above each one of the terms on the right--most side.
As for the first term, we have
\begin{eqnarray}
\label{15}
I(U^N;V^N,K|W^N,Z^n)&=&I(U^N;K|W^N,Z^n)+I(U^N;V^N|K,W^N,Z^n)\nonumber\\
&\le&H(K|W^N,Z^n)+H(V^N|K,W^N,Z^n)-H(V^N|U^N,K,W^N,Z^n)\nonumber\\
&\le&H(K)+H(V^N|W^N)-H(V^N|U^N,W^N)\nonumber\\
&\le&N(R+\epsilon)+NI(U;V|W)\nonumber\\
&=&N[R+I(U;V|W)+\epsilon],
\end{eqnarray}
where in the second inequality we have used the 
fact that $V^N\ominus(U^N,W^N)\ominus(Z^n,K)$ is a Markov chain.
As for the second term on the r.h.s.\ of (\ref{2terms}), we have:
\begin{eqnarray}
\label{3terms}
H(U^N|V^N,K,W^N,Z^n)&=&H(U^N|W^N)-I(U^N;V^N,K,Z^n|W^N)\nonumber\\
&=&NH(U|W)-I(U^N;V^N|W^N)-I(U^N;K,Z^n|W^N,V^N)\nonumber\\
&=&N[H(U|W)-I(U;V|W)]-I(U^N;K,Z^n|W^N,V^N)\nonumber\\
&=&NH(U|V,W)-I(U^N;K,Y^n|W^N,V^N)+\nonumber\\
& &[I(U^N;K,Y^n|W^N,V^N)-I(U^N;K,Z^n|W^N,V^N)].
\end{eqnarray}
We proceed by deriving 
a lower bound to $I(U^N;K,Y^n|W^N,V^N)$ and an upper bound
to the bracketed term in the last expression. As for the former, we have:
\begin{equation}
\label{17}
I(U^N;K,Y^n|W^N,V^N)\ge I(U^N;Y^n|W^N,V^N,K)\ge NR_{U|V}(D+\epsilon),
\end{equation}
where the second inequality has been proven above (compare the right--hand side of the first line
of eq.\ (\ref{2nd}) with the right--most side of that equation).
As for the upper bound
to the bracketed term of the right--most side of (\ref{3terms}), we have:
\begin{eqnarray}
\label{18}
& &I(U^N;K,Y^n|W^N,V^N)-I(U^N;K,Z^n|W^N,V^N)\nonumber\\
&\eqa&I(U^N;K,Y^n,W^N,V^N)-I(U^N;K,Z^n,W^N,V^N)\nonumber\\
&\eqb&I(U^N,K;K,Y^n,W^N,V^N)-I(U^N,K;K,Z^n,W^N,V^N)\nonumber\\
&\eqc&I(U^N,K,X^n;K,Y^n,W^N,V^N)-I(U^N,K,X^n;K,Z^n,W^N,V^N)\nonumber\\
&\eqd&I(U^N,K,X^n;Y^n|K,W^N,V^N)-I(U^N,K,X^n;Z^n|K,W^N,V^N)\nonumber\\
&\eqe&I(U^N,X^n;Y^n|K,W^N,V^N)-I(U^N,X^n;Z^n|K,W^N,V^N)\nonumber\\
&=&I(X^n;Y^n|K,W^N,V^N)-I(X^n;Z^n|K,W^N,V^N)\nonumber\\
&\eqf&H(Y^n|K,W^N,V^N)-H(Z^n|K,W^N,V^N)+\nonumber\\
& &H(Z^n|X^n,K,W^N,V^N)-H(Y^n|X^n,K,W^N,V^N)\nonumber\\
&=&\sum_{i=1}^n[H(Y_i|Y^{i-1},K,W^N,V^N)-H(Z_i|Z^{i-1},K,W^N,V^N)+\nonumber\\
& &H(Z_i|X_i,K,W^N,V^N)-H(Y_i|X_i,K,W^N,V^N)]\nonumber\\
&\leg&\sum_{i=1}^n[H(Y_i|Y^{i-1},K,W^N,V^N)-H(Z_i|Z^{i-1},Y^{i-1},K,W^N,V^N)+\nonumber\\
& &H(Z_i|X_i,K,W^N,V^N)-H(Y_i|X_i,K,W^N,V^N)]\nonumber\\
&\eqh&\sum_{i=1}^n[H(Y_i|Y^{i-1},K,W^N,V^N)-H(Z_i|Y^{i-1},K,W^N,V^N)+\nonumber\\
& &H(Z_i|X_i,Y^{i-1},K,W^N,V^N)-H(Y_i|X_i,Y^{i-1},K,W^N,V^N)]\nonumber\\
&=&\sum_{i=1}^n[I(X_i;Y_i|Y^{i-1},K,W^N,V^N)-I(X_i;Z_i|Y^{i-1},K,W^N,V^N)]\nonumber\\
&=&\sum_{i=1}^n[H(X_i|Z_i,Y^{i-1},K,W^N,V^N)-H(X_i|Y_i,Y^{i-1},K,W^N,V^N)]\nonumber\\
&\eqi&\sum_{i=1}^n[H(X_i|Z_i,Y^{i-1},K,W^N,V^N)-H(X_i|Y_i,Z_i,Y^{i-1},K,W^N,V^N)]\nonumber\\
&=&\sum_{i=1}^nI(X_i;Y_i|Z_i,Y^{i-1},K,W^N,V^N)
\end{eqnarray}
where (a) is by adding and subtracting $I(U^N;V^N,W^N)$,
(b) is by adding $I(K;K,Y^n,W^N,V^N|U^N)=H(K|U^N)$
and subtracting $I(K;K,Z^n,W^N,V^N|U^N)=H(K|U^N)$,
(c) is by the fact that $X^n$ is a function of $U^N$ and $K$,
(d) is by adding and subtracting $I(U^N,K,X^n;K,W^N,V^N)$,
(e) is by the fact that $K$ is degenerate as it appears in the conditioning,
(f) is by the fact that 
$U^N\ominus (X^n,K,W^N,V^N)\ominus Y^n\ominus Z^n$ is a Markov chain,
(g) is because conditioning reduces entropy,
(h) is because $Z^{i-1}\ominus (Y^{i-1},K,W^N,V^N)\ominus Z_i$ and
$Z_i\ominus Y_i\ominus(X_i,K,V^N,W^N)\ominus Y^{i-1}$ are Markov chains,
and (i) is because $X_i\ominus(Y_i,Y^{i-1},K,W^N,V^N)\ominus Z_i$ is a Markov chain.

At this point, we are after an upper 
bound to $\sum_{i=1}^nI(X_i;Y_i|Z_i,Y^{i-1},K,W^N,V^N)$,
subject to the fact that 
\begin{eqnarray}
\label{arg}
NR_{U|V}(D+\epsilon)&\le& I(X^n;Y^n|K,V^N,W^N)\nonumber\\
&=&\sum_{i=1}^n
[H(Y_i|Y^{i-1},K,V^N,W^N)-H(Y_i|X_i,Y^{i-1},K,V^N,W^N)]\nonumber\\
&=&\sum_{i=1}^nI(X_i;Y_i|Y^{i-1},K,V^N,W^N)
\end{eqnarray}
where, once again, the first inequality has been proved already
in (\ref{2nd}).
For given $k$,$v^N$, $w^N$, and $y^{i-1}$, $i=1,2,\ldots,n$, let
\begin{equation}
\alpha_i(k,v^N,w^N,y^{i-1})=I(X_i;Y_i|K=k,V^N=v^N,W^N=w^N,Y^{i-1}=y^{i-1})
\end{equation}
and
\begin{equation}
\beta_i(k,v^N,w^N,y^{i-1})=\bE\{\phi(X_i)|K=k,V^N=v^N,W^N=w^N,Y^{i-1}=y^{i-1}\}.
\end{equation}
Obviously, by definition of the function $\Gamma$,
\begin{eqnarray}
& &I(X_i;Y_i|Z_i,K=k,V^N=v^N,W^N=w^N,Y^{i-1}=y^{i-1})\nonumber\\
&\le&\Gamma(\alpha_i(k,v^N,w^N,y^{i-1}),\beta_i(k,v^N,w^N,y^{i-1})).
\end{eqnarray}
Thus,
\begin{eqnarray}
\label{22}
& &\sum_{i=1}^nI(X_i;Y_i|Z_i,Y^{i-1},K,W^N,V^N)\nonumber\\
&=&\sum_{i=1}^n\sum_{k,v^N,w^N,y^{i-1}}
\mbox{Pr}\{K=k,V^N=v^N,W^N=w^N,Y^{i-1}=y^{i-1}\}\times\nonumber\\
& &I(X_i;Y_i|Z_i,K=k,V^N=v^N,W^N=w^N,Y^{i-1}=y^{i-1})\nonumber\\
&\le&\sum_{i=1}^n\sum_{k,v^N,w^N,y^{i-1}}
\mbox{Pr}\{K=k,V^N=v^N,W^N=w^N,Y^{i-1}=y^{i-1}\}\times\nonumber\\
& &\Gamma(\alpha_i(k,v^N,w^N,y^{i-1}),\beta_i(k,v^N,w^N,y^{i-1}))\nonumber\\
&\lea&n\Gamma\left(\frac{1}{n}\sum_{i=1}^n\sum_{k,v^N,w^N,y^{i-1}}
\mbox{Pr}\{K=k,V^N=v^N,W^N=w^N,Y^{i-1}=y^{i-1}\}\cdot\alpha_i(k,v^N,w^N,y^{i-1}),\right.\nonumber\\
& &\left.\frac{1}{n}\sum_{i=1}^n\sum_{k,v^N,w^N,y^{i-1}}
\mbox{Pr}\{K=k,V^N=v^N,W^N=w^N,Y^{i-1}=y^{i-1}\}\cdot\beta_i(k,v^N,w^N,y^{i-1})
\right)\nonumber\\
&=&n\Gamma\left(\frac{1}{n}\sum_{i=1}^nI(X_i;Y_i|Y^{i-1},K,W^N,V^N),\frac{1}{n}\sum_{i=1}^n\bE\{\phi(X_i)\}
\right)\nonumber\\
&\leb&n\Gamma\left(\frac{N}{n}\cdot R_{U|V}(D+\epsilon),Q+\epsilon\right)\nonumber\\
&\lec&n\Gamma\left(\frac{R_{U|V}(D+\epsilon)}{\lambda+\epsilon},Q+\epsilon\right)\nonumber\\
&\led&N(\lambda+\epsilon)\Gamma\left(\frac{R_{U|V}(D+\epsilon)}{\lambda+\epsilon},Q+\epsilon\right),
\end{eqnarray}
where (a) follows from the concavity of $\Gamma(r,q)$ jointly 
in both arguments,\footnote{This can readily be
verified as a trivial extension of \cite[Lemma 1]{Wyner75} which accounts for the generalized power constraint.}
together with its non--increasing monotonicity in $r$ and non--decreasing monotonicity in $q$,
(b)-- from (\ref{arg}) and the non--increasing 
monotonicity of the function $\Gamma(\cdot)$, and (c) and (d) --
from the postulate that the bandwidth 
expansion factor of the codec does not
exceed $\lambda+\epsilon$.
Combining eqs.\ (\ref{1stbound}), (\ref{2terms}), (\ref{15}), 
(\ref{3terms}), (\ref{17}), (\ref{18}), and (\ref{22}),
and using the arbitrariness of $\epsilon$ with continuity
considerations, we get 
\begin{eqnarray}
\Delta&\le& 
\min\left\{H(U|W),R+I(U;V|W)+H(U|V,W)+
\lambda\Gamma\left(\frac{R_{U|V}(D)}{\lambda},Q\right)-R_{U|V}(D)\right\}\nonumber\\
&=&\min\left\{H(U|W),H(U|W)+R+
\lambda\Gamma\left(\frac{R_{U|V}(D)}{\lambda},Q\right)-R_{U|V}(D)\right\}\nonumber\\
&=&H(U|W)-\left[R_{U|V}(D)-\lambda\Gamma\left(\frac{R_{U|V}(D)}{\lambda},Q\right)-R\right]_+\nonumber\\
&=&\Delta^*(\lambda,R,D,Q),
\end{eqnarray}
which establishes the
converse part of Theorem 1.

\section{Proof of the Direct Part of Theorem 1}

We begin with the following
chain of equalities and inequalities:
\begin{eqnarray}
\label{genlowerbound}
N\Delta&=&H(U^N|W^N,Z^n)\nonumber\\
&=&H(U^N,Z^n|W^N)-H(Z^n|W^N)\nonumber\\
&=&H(U^N,Z^n,X^n,K|W^N)-H(X^n,K|U^N,Z^n,W^N)-H(Z^n|W^N)\nonumber\\
&=&H(X^n,K,U^N|W^N)+H(Z^n|X^n,K,U^N,W^N)-\nonumber\\
& &H(X^n,K|U^N,Z^n,W^N)-H(Z^n|W^N)\nonumber\\
&=&H(Z^n|X^n,K,U^N,W^N)+H(U^N|W^N)+\nonumber\\
& &[H(X^n,K|U^N,W^N)-H(X^n,K|U^N,Z^n,W^N)]-H(Z^n|W^N)\nonumber\\
&=&NH(U|W)+H(Z^n|X^n,K,U^N,W^N)+\nonumber\\
& &I(X^n,K;Z^n|U^N,W^N)-H(Z^n|W^N)\nonumber\\
&\gea&NH(U|W)+H(Z^n|X^n)+\nonumber\\
& &I(X^n,K;Z^n|U^N,W^N)-H(Z^n)\nonumber\\
&=&NH(U|W)+I(X^n,K;Z^n|U^N,W^N)-I(X^n;Z^n)\nonumber\\
&=&NH(U|W)+I(K;Z^n|U^N,W^N)+I(X^n;Z^n|U^N,W^N,K)-I(X^n;Z^n)\nonumber\\
&=&NH(U|W)+H(K|U^N,W^N)-H(K|U^N,W^N,Z^n)+\nonumber\\
& &I(X^n;Z^n|U^N,W^N,K)-I(X^n;Z^n)\nonumber\\
&\eqb&NH(U|W)+H(K)-H(K|U^N,W^N,Z^n)+\nonumber\\ 
& &I(X^n;Z^n|U^N,W^N,K)-I(X^n;Z^n)\nonumber\\
&\eqc&NH(U|W)+NR-H(K|U^N,W^N,Z^n)+\nonumber\\
& &I(X^n;Z^n|U^N,W^N,K)-I(X^n;Z^n),
\end{eqnarray}
where (a) follows from the fact that 
$(K,U^N,W^N)\ominus X^n\ominus Z^n$ is a Markov chain,
(b) -- from the fact that $K$ is 
independent of $(U^N,W^N)$, and (c) -- by assuming that $H(K)=NR$.
While this chain of equalities and inequalities holds for any codec,
then in order to proceed, we will 
have to be specific, from now on, about the structure
and the properties of the codec. In particular, 
referring to the right--most side of the above lower bound
to $N\Delta$, then in order
to prove the direct part, we will have to prove that
for our proposed codec (and as long as $R$ is not too large): (i)
$H(K|U^N,W^N,Z^n)$ is small, (ii) $I(X^n;Z^n)$ is essentially smaller
than $nI(X;Z)$, and (iii) $I(X^n;Z^n|U^N,W^N,K)$ is essentially larger
$[nI(X;Y)-NR_{U|V}(D)]$, 
where in (ii) and (iii) the distribution of the random variable $X$ is the
achiever of $\Gamma(R_{U|V}(D)/\lambda,Q)$.

Fix an arbitrarily small $\epsilon > 0$, and 
let $D$ satisfy $R_{U|V}(D)\le \lambda C_{X\to Y}-\epsilon$, where $C_{X\to Y}$ is the
capacity of the channel $P_{Y|X}$.
Given such $D$ and $\epsilon > 0$, let $X^*$ denote
the channel input variable
that achieves 
$\Gamma((R_{U|V}(D)+\epsilon)/\lambda,Q)$. Let $Y^*$ and $Z^*$
denote the channel output variables 
induced by $X^*$ and the channels $P_{Y|X}$ and $P_{Z|Y}$,
respectively. Thus, 
\begin{equation}
I(X^*;Y^*)-I(X^*;Z^*)=\Gamma\left(\frac{R_{U|V}(D)+\epsilon}{\lambda},Q\right)
\end{equation}
and
\begin{equation}
\label{constraint}
I(X^*;Y^*)\ge \frac{R_{U|V}(D)+\epsilon}{\lambda}.
\end{equation}
Let us further suppose now that for the resulting optimal RV's $X^*$, $Y^*$, and $Z^*$, we have:
\begin{equation}
\label{pos}
\frac{R_{U|V}(D)}{\lambda} > I(X^*;Y^*)-I(X^*;Z^*).
\end{equation}
In the sequel, we will handle separately the case where (\ref{pos}) does not hold.
Further, let $\calT_n$ denote the set of $n^{-1/4}$--typical $n$--sequences
with components in $\calX$, i.e., the set of sequences for which
the relative frequency of each $x\in\calX$
differs from $P_{X^*}(x)$ by no more than $n^{-1/4}$.
The following lemma, which is Lemma 8 
of \cite{Wyner75}, guarantees that if the encoder
is such that, with high probability $X^n\in \calT_n$, then
condition (ii) above is essentially satisfied:

\begin{lemma}\cite[Lemma 8]{Wyner75}
Let $X^n$ and $Z^n$ be induced by an aribtrary encoder
and the cascaded channel from $X^n$ to $Z^n$:
\begin{equation}
\label{condition2}
\frac{I(X^n;Z^n)}{n}\le I(X^*;Z^*)+\mbox{Pr}
\{X^n\in \calT_n^c\}\cdot \log|\calX|+f_1(n),
\end{equation}
where $f_1(n)\to 0$ as $n\to\infty$.
\end{lemma}

\begin{figure}[ht]
\hspace*{2cm}\input{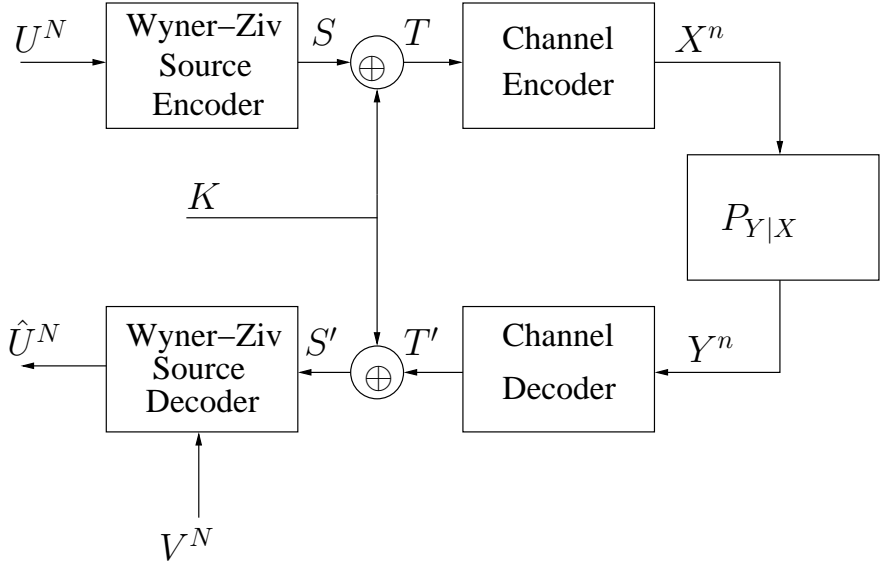}
\caption{Encoder and decoder for the direct part.}
\label{sys}
\end{figure}

Note that whenever $X^n\in \calT_n$, the generalized power constraint is also essentially satisfied.
It remains to handle conditions (i) and (iii).
Consider next the encoder and the decoder of the
legitimate receiver, depicted in Fig.\ \ref{sys}. 
The source vector $U^N$ is first
compressed by a Wyner--Ziv encoder, designed for distortion level $D$ and 
side information $V^N$, to a string of bits, $S=F_E(U^N)$,
whose length does not exceed $N[R_{U|V}(D)+\epsilon/2]\le n[I(X^*;Y^*)-\epsilon/(2\lambda)]$. 
Now, let us select $R$ in the range
\begin{equation}
\label{keyrate}
0\le R < R_{U|V}(D)-\lambda[I(X^*;Y^*)-I(X^*;Z^*)]-\epsilon,
\end{equation}
where the right--most side is positive due to (\ref{pos}).
The key $K$ is a string of $NR$ 
purely random bits, which are XORed with (the first) $NR$ bits of $S$ (one time pad). The resulting (partially)
encrypted bit string, $T$, which will be 
represented by $T=S\oplus K=F_E(U^N)\oplus K$ (although it is possible 
that only some of the bits of $S$ are XORed with those of $K$),
is the message to be conveyed across the channel. Now, let
\begin{equation}
q_t\dfn \mbox{Pr}\{T=t\}, ~~~~t=1,2,\ldots, M\dfn 2^{N[R_{U|V}(D)+\epsilon/2]}.
\end{equation}
Next, let $M_1=M_2M$, where $M_2$ is a positive integer to be specified in the sequel.
Let $\{\bx_m\}_{m=1}^{M_1}$ be a subset of $\calX^n$, which can be viewed as a code for the channel
$P_{Y|X}$ or $P_{Z|X}$. The channel encoder and decoder in Fig.\ \ref{sys} work as follows.
They both share a partition of $\{\bx_m\}_{m=1}^{M_1}$ into $M$ sub--codes, $\calC_1,\calC_2,\ldots,\calC_M$,
each of size $M_2$. Let $\calC_t=\{\bx_{(i-1)M_2+1},\ldots,\bx_{iM_2}\}$, $t=1,2,\ldots,M$.
When $T=t$, the 
channel encoder outputs a vector $X^n$ which is a (uniformly) randomly
chosen member of sub--code $\calC_t$. 
Thus, for $t=1,2,\ldots,M$, $\tau=1,2,\ldots,M_2$,
\begin{equation}
\label{condprior}
\mbox{Pr}\{X^n=\bx_{(t-1)M_2+\tau}|T=t\}=\frac{1}{M_2}
\end{equation}
and
\begin{equation}
\label{prior}
\mbox{Pr}\{X^n=\bx_{(t-1)M_2+\tau}\}=\frac{q_t}{M_2}.
\end{equation}
As mentioned earlier, the set $\{\bx_m\}_{m=1}^{M_1}$
can be thought of as a code for the channel 
$P_{Y|X}$, where the prior probabilities of the codewords are
given by (\ref{prior}). Let $T'=G(Y^n)$ denote the Bayes--optimal 
decoder for this code and these prior probabilities, 
which estimates the index $t$ of the
sub--code $\calC_t$ that contains the transmitted codeword $X^n$.
Let $\delta=\delta_Y(\bx_1,\ldots,\bx_{M_1})\dfn
\mbox{Pr}\{T'\ne T\}$. 
Obviously, if $\delta$ is small, namely, if $T'=T$ with
high probability, then the Wyner--Ziv decoder 
$\hat{U}^N=F_D(S',V^N)=F_D(T'\oplus K,V^N)$ would output
the ``correct'' reconstruction vector 
within distortion $D$, with the same probability.

Next, observe that each sub--code $\calC_t$ may serve as a channel code for the degraded
channel $P_{Z|X}$, provided that the 
corresponding decoder is informed of $t$. Let $\delta_t=\delta_Z(\calC_t)$,
$t=1,2,\ldots,M$, denote the error probability of code 
$\calC_t$ w.r.t.\ the channel $P_{Z|X}$
when the decoder that observes $Z^n$ is informed of $t$. 
Finally, let $\bar{\delta}=\sum_{t=1}^Mq_t\delta_t$. 
With these definitions, we next make our first step to handle condition (iii).

Let $U^N$ and $K$ be such that $T=F_E(U^N)\oplus K=t$. 
Then, the channel input, given $T=t$,
is distributed according to (\ref{condprior}), that is, 
$X^n$ is a randomly chosen member of $\calC_t$,
Thus, $H(X^n|T=t)=\log M_2$.
Since $\delta_t$ is the probability of 
error associated with $\calC_t$, Fano's inequality yields:
\begin{equation}
H(X^n|Z^n,T=t)\le h(\delta_t)+\delta_t\log M_2
\le 1+\delta_t\log M_2
\end{equation}
where $h(\cdot)$ is the
binary entropy function $h(a)=-a\log a-(1-a)\log(1-a)$.
It follows then that
\begin{equation}
I(X^n;Z^n|T=t)\ge (1-\delta_t)\log M_2-1,
\end{equation}
which upon averaging over $\{t\}$ with weights $\{q_t\}$, yields
\begin{equation}
I(X^n;Z^n|T)\ge (1-\bar{\delta})\log M_2-1.
\end{equation}
On the other hand,
\begin{eqnarray}
I(X^n;Z^n|T)&=&H(Z^n|T)-H(Z^n|X^n,T)\nonumber\\
&=&H(Z^n|T,U^N,W^N,K)-H(Z^n|X^n,T,U^N,W^N,K)\nonumber\\
&\le&H(Z^n|U^N,W^N,K)-H(Z^n|X^n,U^N,W^N,K)\nonumber\\
&=&I(X^n;Z^n|U^N,W^N,K)
\end{eqnarray}
where the second equality 
is due to the Markov relation $(U^N,W^N,K)\ominus T\ominus X^n\ominus Z^n$.
Thus, we have established the inequality
\begin{equation}
\label{condition3}
I(X^n;Z^n|U^N,W^N,K)\ge (1-\bar{\delta})\log M_2-1.
\end{equation}
In the sequel, we will choose $M_2$ so as to meet condition (iii).

We next move on to handle condition (i).
For every $S=s$, let $\calC_s'$ denote the union
of all $2^{NR}$ codebooks $\{\calC_{t}, t=s\oplus k\}_{k=0}^{2^{NR}-1}$, 
and let $\delta_s'=\delta_Z(\calC_s')$ denote
the error probability of $\calC_s'$ w.r.t.\ the 
channel $P_{Z|X}$ when the decoder is informed of $s$. Let
\begin{equation}
p_s\dfn \mbox{Pr}\{S=s\}, ~~~~s=1,2,\ldots, M.
\end{equation}
Finally, let $\bar{\delta}'=\sum_sp_s\delta_s'$.
With these definitions, let us now derive an upper bound on $H(K|U^N,W^N,Z^n)$:
\begin{eqnarray}
\label{condition1}
H(K|U^N,W^N,Z^n)&\le&H(K|U^N,F_E(U^N),Z^n)\nonumber\\
&=&H(K|U^N,S,Z^n)\nonumber\\
&\le&H(S\oplus T|S,Z^n)\nonumber\\
&=&H(T|S,Z^n)\nonumber\\
&=&\sum_{s=1}^Mp_sH(T|S=s,Z^n)\nonumber\\
&\le&\sum_{s=1}^Mp_s[h(\delta_s')+\delta_s'\log(2^{NR}M_2)]\nonumber\\
&\le&1+\bar{\delta}'(NR+\log M_2),
\end{eqnarray}
where the third inequality is again Fano's inequality, and where
we have also used the fact that $\delta_s'$ 
is an upper bound of the probability
of error in estimating $T$, since $T$ is only the index 
$t$ of the codebook $\calC_t$ to
which the estimated codeword belongs.

To summarize our findings thus far, we 
substitute eqs.\ (\ref{condition1}), 
(\ref{condition2}) and (\ref{condition3}) into
(\ref{genlowerbound}), divide by $N$, and get:
\begin{eqnarray}
\label{2ndlb}
\Delta&\ge&H(U|W)+R+
(1-\bar{\delta})\cdot\frac{\log M_2}{N}-\frac{2}{N}-
\bar{\delta}'\left(R+\frac{\log M_2}{N}\right)-\nonumber\\
& &\lambda[I(X^*;Z^*)+\mbox{Pr}
\{X^n\in \calT_n^c\}\cdot\log|\calX|+f_1(n)].
\end{eqnarray}
Now, let us select
\begin{equation}
M_1=2^{n[I(X^*;Y^*)-\epsilon/(2\lambda)]}
\end{equation}
and
\begin{equation}
M_2=\frac{M_1}{M}=2^{n[I(X^*;Y^*)-R_{U|V}(D)/\lambda-\epsilon/\lambda]}.
\end{equation}
Applying this to (\ref{2ndlb}), we get
\begin{eqnarray}
\label{almostdone}
\Delta&\ge&H(U|W)+R+
(1-\bar{\delta})[\lambda I(X^*;Y^*)-R_{U|V}(D)-\epsilon]-\frac{2}{N}-
\bar{\delta}'(R+\log |\calX|)-\nonumber\\
& &\lambda[I(X^*;Z^*)+\mbox{Pr}
\{X^n\in \calT_n^c\}\cdot\log|\calX|+f_1(n)]\nonumber\\
&\ge&H(U|W)+R+\lambda[I(X^*;Y^*)-I(X^*;Z^*)]-R_{U|V}(D)-\nonumber\\
& &\left\{\epsilon+\frac{2}{N}+
(\bar{\delta}+\bar{\delta}')(R+\log |\calX|)+\lambda[\mbox{Pr}
\{X^n\in \calT_n^c\}\cdot\log|\calX|+f_1(n)]\right\}\nonumber\\
&=&H(U|W)+R+\lambda\Gamma\left(\frac{R_{U|V}(D)+\epsilon}{\lambda},Q\right)-R_{U|V}(D)-\nonumber\\
& &\left\{\epsilon+\frac{2}{N}+(
\bar{\delta}+\bar{\delta}')(R+\log |\calX|)+\lambda[\mbox{Pr}
\{X^n\in \calT_n^c\}\cdot\log|\calX|+f_1(n)]\right\}.
\end{eqnarray}
Finally, to prove that the expected 
distortion of $\hat{U}^N$ relative to $U^N$ is essentially $D$,
and to prove that $\Delta$ essentially meets 
the upper bound $\Delta^*(\lambda,R,D,Q)$
(namely, that the last term on the right--most side of (\ref{almostdone}) is arbitrarily small 
for large $N$), we have to prove the existence of
a code $\{\bx_m\}_{m=1}^{M_1}$ for which $\delta$, $\bar{\delta}$, 
$\bar{\delta}'$ and $\mbox{Pr}\{X^n\in \calT_n^c\}$
are all simultaneously arbitrarily small for large $N$.

To this end, let us define $\mu(\bx)=1\{\bx\in \calT_n^*\}$, and for a given code $\{\bx_m\}_{m=1}^{M_1}$, 
let $\delta_Y^m(\bx_1,\ldots,\bx_{M_1})$ 
denote the error probability w.r.t.\ the channel $P_{Y|X}$
with prior probabilities 
$\{q_t\}$ as given in (\ref{prior}), when $\bx_m$ is transmitted. Then,
$$\delta=\sum_{s=0}^{M-1}p_s\sum_{k=0}^{2^{NR}-1}\frac{1}{2^{NR}}
\sum_{m=(s\oplus k)M_2+1}^{(s\oplus k+1)M_2} 
\frac{1}{M_2}\delta_Y^m(\bx_1,\ldots,\bx_{M_1}).$$
Further, let $\delta_t=\delta_Z(\calC_t)$, $\delta_s'=\delta_Z(\calC_s')$, 
$\bar{\delta}$, 
and $\bar{\delta}'$ be defined 
as above. Then,
\begin{eqnarray}
\Phi(\bx_1,\ldots,\bx_{M_1})&\dfn&
\mbox{Pr}\{X^n\in \calT_n^c\}+\delta+\bar{\delta}+\bar{\delta}'\nonumber\\
&=&\sum_{s=0}^{M-1}p_s\sum_{k=0}^{2^{NR}-1}
\frac{1}{2^{NR}}\sum_{m=(s\oplus k)M_2+1}^{(s\oplus k+1)M_2} \frac{1}{M_2}
[\mu(\bx_m)+\delta_Y^m(\bx_1,\ldots,\bx_{M_1})+\nonumber\\
& &\delta_Z(\calC_{s\oplus k})
+\delta_Z(\calC_s')].
\end{eqnarray}
Now, suppose that $\{\bx_m\}_{m=1}^{M_1}$ are selected at random,
with each $\bx_m$ chosen independently according to
$P_{X^*}^n(x^n)=\prod_{j=1}^nP_{X^*}(x_j)$. To prove that there
exists a sequence of codes for 
which $\Phi\to 0$ as $n\to\infty$, all we have to
show is that $\bE\Phi\to 0$. But 
$$\bE\Phi=\bE\mu(\bX)+\bE\delta_Y^m(\bX_1,\ldots,\bX_{M_1})
+\bE\delta_Z(\calC_{s\oplus k})+\bE\delta_Z(\calC_s'),$$
where the indices $s$, $k$, and $m\in\{(s\oplus k)M_2+1,\ldots,(s\oplus k +1)
M_2\}$ are now immaterial.
The first term tends to zero by the weak law of large numbers.
The second term tends to zero by the ordinary random channel coding argument
as the rate of the code
$\{\bx_m\}_{m=1}^{M_1}$ is less than $I(X^*;Y^*)$ (cf.\ the choice
of $M_1$ above).
By the same token,
the fourth term vanishes with $n$, as $\calC_s'$ is a random code 
of size $2^{NR}M_2$, and so its 
rate (cf.\ (\ref{keyrate})) is 
$$R/\lambda+I(X^*;Y^*)-R_{U|V}(D)/\lambda < I(X^*;Z^*)-\epsilon/\lambda,$$
which means that it is reliable for the channel $P_{Z|X}$ on the average.
A--fortiori, the third term decays with $n$ as $\calC_t$
is even a smaller random code. By a simple application of the Chebychev inequality,
with probability of at least $\frac{2}{3}$, the random of selection of the code
yields $\Phi(\bX_1,\ldots,\bX_{M_1})\le 3\bE\Phi$, which is still vanishingly small. On the other hand,
since the codeword components are selected i.i.d.\ under $P_{X^*}$, then by the weak law of large numbers,
for every $\epsilon > 0$ and large enough $n$ and $M_1$, we have, 
with probability that tends to unity, and in particular, larger than $\frac{1}{2}$ from some point on:
\begin{equation}
\label{power}
\sum_{t=1}^M\sum_{\tau=1}^{M_2}\frac{q_t}{M_2}\cdot\frac{1}{n}\sum_{j=1}^n
\phi([\bX_{(t-1)M_2+\tau}](j)) \le Q+\epsilon
\end{equation}
where $[\bX_{(t-1)M_2+\tau}](j)$ is the $j$--th component of the codeword $\bX_{(t-1)M_2+\tau}$.
Since $\frac{2}{3}+\frac{1}{2} > 1$, it follows then that there exist codes for
which both $\Phi(\bx_1,\ldots,\bx_{M_1})\le 3\bE\Phi$ (and hence all components of $\Phi$ must be small) and
the power constraint (\ref{power}) holds at the same time.

The maximum secrecy of $H(U|W)$
is, of course, approached by letting $R$ be arbitrarily
close to (but strictly smaller than) 
$R_{U|V}(D)-\lambda\Gamma(R_{U|V}(D)/\lambda)$.

Finally, for completeness, we give a sketchy 
description of how the proof of the direct part should
be slighlty modified in the (simpler) case where eq.\ (\ref{pos}) does not hold, namely,
\begin{equation}
\label{neg}
\frac{R_{U|V}(D)}{\lambda} \le I(X^*;Y^*)-I(X^*;Z^*).
\end{equation}
Note that in this case, the achievable upper bound on $\Delta$, asserted in
Theorem 1, becomes $H(U|W)$ even for $R=0$, as the bracketed term therein is non--positive.
In the case, we will not use the key at all, i.e., $R=0$ and $K$ is degenerate.
Thus, (\ref{genlowerbound}) becomes now:
$$N\Delta\ge NH(U|W)+I(X^n;Z^n|U^N,W^N)-I(X^n;Z^n).$$
As before, $I(X^n;Z^n)$ is essentially 
upper bounded by $nI(X^*;Z^*)$ using Lemma 1, and so,
we only have to deal with the term $I(X^n;Z^n|U^N,W^N)$ and show that it is essentially 
lower bounded by $nI(X^*;Z^*)$. To this end, let us re--define $M_1$ as
$$M_1=2^{NR_{U|V}(D)+n[I(X^*;Z^*)-\epsilon/(2\lambda)]},$$
and $M$ as before, so,
$$M_2=\frac{M_1}{M}= 2^{n[I(X^*;Z^*)-\epsilon/\lambda]}.$$
Now, since $NR_{U|V}(D)+nI(X^*;Z^*) < nI(X^*;Y^*)$ 
(cf.\ (\ref{neg})), the full codeword $\bx_m$ can be reliably decoded
at the legitimate decoder, as before. 
Also, since each sub--code $\calC_t$ is, again, of rate less than $I(X^*;Z^*)$,
then it can be decoded reliably by 
the wiretapper, provided that s/he is informed of $t$,
thus $I(X^n;Z^n|U^N,W^N)$ is again, essentially 
lower bounded by $\log M_2=n[I(X^*;Z^*)-\epsilon/\lambda]$.
This completes the proof of direct part of Theorem 1.

%\section*{Acknowledgements}
%\section*{Appendix}
%\renewcommand{\theequation}{A.\arabic{equation}}
%    \setcounter{equation}{0}

%\clearpage

\end{document}